# Generic Short-Time Propagation of Sharp-Boundaries Wave Packets

Er'el Granot[(1)] and Avi Marchewka[(2)]

[(1)] *Department of Electrical and Electronics Engineering, College of Judea and Samaria, Ariel, Israel*
[(2)]*Kibbutzim College of Education, Ramat-Aviv, 104 Namir Road 69978 Tel-Aviv, Israel*

**Abstract**

A general solution to the "shutter" problem is presented. The propagation of an arbitrary initially bounded wavefunction is investigated, and the general solution for *any* such function is formulated. It is shown that the *exact* solution can be written as an expression that depends only on the values of the function (and its derivatives) at the boundaries. In particular, it is shown that at short times ($t \ll 2mx^2/\hbar$, where $x$ is the distance to the boundaries) the wavefunction propagation depends *only* on the wavefunction's values (or its derivatives) at the boundaries of the region. Finally, we generalize these findings to a non-singular wavefunction (i.e., for wavepackets with finite-width boundaries) and suggest an experimental verification.

**I. Introduction**. In 1952 Moshinsky was the first to address the "Shutter problem"[1]. He formulated the following problem: a monochromatic beam of particles is interrupted at a certain place by a shutter, which can move perpendicularly to the beam. Suppose at a certain time the shutter is opened abruptly, what will be the particles' probability density measurement a distance $x$ from the shutter? To address this question he solved the Schrödinger equation with the *singular* initial condition

$$\varphi(x, t=0) = \exp(ikx)\Theta(-x) \qquad (1)$$

(where $\Theta(x) = \{1 \text{ for } x \geq 0, \ 0 \text{ for } x < 0\}$ is the step function and $k$ is the particles' wavenumber).

He found that unlike in classical mechanics, where the beam's current jumps suddenly from zero to the stationary value at $t = x/v = xm/\hbar k$, in Quantum Mechanics there is a diffraction effect in time. He also mentioned the resemblance to the Sommefeld's theory of diffraction.

This function includes singularities in time and space, and therefore his derivations can naturally raise skepticism about their relevance to real experiments.

It took more than four decades until this problem could be addressed experimentally and Moshinsky's derivation was validated [2].

The recent development in femto-second lasers [3] optical tweezers and quantum trapping [4, 5] makes it possible to localize quantum particles in space, and since the trap is actually made of light, they can be released instantaneously. Moreover, the ability to modify the laser trapping-beam makes it possible to control the boundaries' shape as well as to sculpture the initial wavepacket[6,7].

This progress in experimental technology naturally raises the motivation for generalizing the shutter problem to an initially localized (in both directions) wavefunction with an *arbitrary* shape.

Recently, the shutter problem was recruited by many researchers in studies of fundamental issues in Quantum Mechanics, like the passage-time in tunneling [8], particle absorption and imaginary potential [9] and quantum measurements and collapse [10].

In this paper we investigate both analytically and numerically the short time expansion of the "shutter problem". First we present a *general solution* to the problem, i.e., the initial condition does not have to be a truncated plane wave[1] nor does it have to be an eigen-bound state of an infinite well [11]. We solve the problem for *any* analytic function that is bounded in a certain region. Then we focus on the short time domain and show a universal phenomenon: at very short times after the constraints are loose the wavefunction far from the initial location region depends *only* on its initial value (and derivatives) at the region's boundaries. We then generalize the conclusions to the general continuous boundaries problem.

**II Theory.** Consider the propagation of a quantum wavefunction, which is analytic except for one singular point (without loss of generality we choose $x = 0$), beyond which it vanishes (see fig.1).

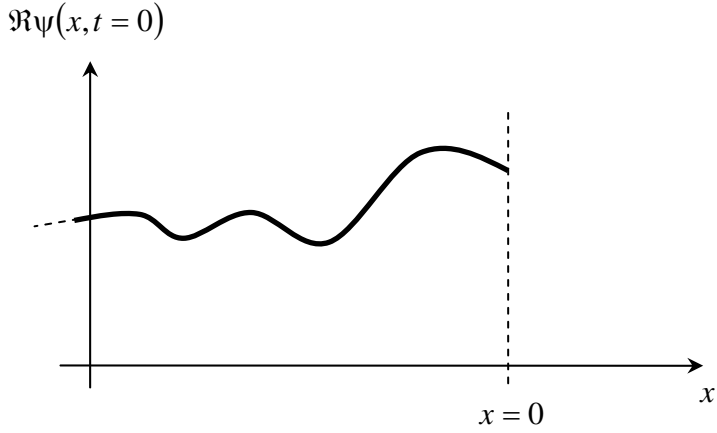

Figure 1: The real part of the wavefunction at $t = 0$ with one singular point.

This problem can be regarded as a generalization of the Moshinsky problem[1,9,11] since any such wave function can be presented as a superposition of truncated plane waves like eq.1.

The solution to the Schrödinger equation with the initial condition (1) can easily be solved with the Green function

$$\varphi(x, t \geq 0) = \int_{-\infty}^{0} \frac{dy}{2\sqrt{\pi t}} \exp\left[i(x-y)^2 / 4t\right] \varphi(y, t = 0) \tag{2}$$

Hereinafter, unless stated otherwise, we choose the dimensionless units $\hbar = 2m = 1$.
The result of this integral is known as the Moshinsky function [12]:

$$\varphi(x, t \geq 0) = \frac{1}{2} \exp(ikx - ik^2 t) \operatorname{erfc}\left[\frac{(x - 2kt)}{2(it)^{1/2}}\right]. \tag{3}$$

In general, for any wave packet $\psi(x,t)$ that vanishes at $x > 0$, with the initial condition $\psi(x, t = 0)$, the solution for any $t \geq 0$ can be directly solved

$$\psi(x, t \geq 0) = \frac{1}{2} \int dk g(k) \exp(ikx - ik^2 t) \operatorname{erfc}\left[\frac{(x - 2kt)}{2(it)^{1/2}}\right] \tag{4}$$

where $g(k) \equiv (2\pi)^{-1} \int dx \psi(x, t = 0) \exp(-ikx)$ is the Fourier counterpart of the initial $\psi(x, t = 0)$.
It is easier to express the solution with the function (see ref.13)
$w(z) \equiv \exp(-z^2) \operatorname{erfc}(-iz)$.

Then, eq. (4) can be rewritten

$$\psi(x, t \geq 0) = \frac{\exp(ix^2 / 4t)}{2} \int w\left[\frac{2tk - x}{2(-it)^{1/2}}\right] g(k) dk \tag{5}$$

The function $w(z)$ can be expanded with respect to $k$, and the integral over this expansion can be expressed as an infinite series of the derivatives of the wavefunction at the boundary $x=0$ at $t=0$.

Thus, the general solution (for every $x$) can be expressed in terms of the functions' derivatives at $x=0$

$$\psi(x,t\geq 0)=\frac{1}{2}\exp\left(i\frac{x^2}{4t}\right)w\left(\frac{-i2t\frac{\partial}{\partial y}-x}{2(-it)^{1/2}}\right)\psi(y,0)\bigg|_{y=0} \quad (6)$$

That is, the information required for the wavefunction propagation in time is encapsulated in the boundary.

Eq. 6 should be understood as the operator expansion [13]

$$\psi(x,t\geq 0)=$$
$$\frac{1}{\sqrt{2\pi}}\exp\left(i\frac{x^2}{4t}\right)\frac{\sqrt{i2t}}{x+2it\frac{\partial}{\partial y}}\left\{1+\sum_{m=1}^{\infty}(-1)^m 1\cdot 3\cdots(2m-1)\left[\frac{2it}{(x+2it\partial/\partial y)^2}\right]^m\right\}\psi(y,0)\bigg|_{y=0} \quad (7)$$

For short times (or for long distances $x$), eq.7 can be expressed as powers series of $t^{1/2}/x$ (if we use real physical units, then the short time requirement is equivalent to $(\hbar/2m)(t/x^2)\ll 1$)

$$\psi(x,t\geq 0)=$$
$$\sqrt{\frac{i}{\pi}}\exp\left(i\frac{x^2}{4t}\right)\left\{\frac{t^{1/2}}{x}-2i\frac{t^{3/2}}{x^3}\left(1+x\frac{\partial}{\partial y}\right)-4\frac{t^{5/2}}{x^5}\left[3\left(1+x\frac{\partial}{\partial y}\right)+x^2\frac{\partial^2}{\partial y^2}\right]\cdots\right\}\psi(y,0)\bigg|_{y=0} \quad (8)$$

Therefore, at very short times (or for long distances $x$) the wavefunction over the entire space depends only on the value of the initial wavefunction at the boundary (the singular) point:

$$\psi(x,0\leq t\ll x^2)\sim\sqrt{\frac{i}{\pi}}\exp\left(i\frac{x^2}{4t}\right)\frac{t^{1/2}}{x}\psi(0,0) \quad (9)$$

It should be noted that the wavefunction increases like $t^{1/2}$ instead of $\sim t$ as in analytic cases.
The dependence on the higher derivatives is negligible. Only when $\psi(0,0)=0$ the next order becomes dominant and the first derivative has an effect. In this case, the first order is

$$\psi(x,t\geq 0)\sim -2\sqrt{\frac{-i}{\pi}}\exp\left(i\frac{x^2}{4t}\right)\frac{t^{3/2}}{x^2}\frac{\partial\psi(y,0)}{\partial y}\bigg|_{y=0}. \quad (10)$$

The oscillating factor $\exp(ix^2/4t)$ does not appear in the probability density $|\psi|^2$; however, it has a *large* influence when there is more than one singular point. A case of special interest is a bounded wavefunction, i.e., a wavefunction that vanishes outside a specific region $a\leq x\leq b$ (where $a$ and $b$ are the left and right boundaries respectively). In this case, there are two singular points (see fig.2). The eigenbound states of an infinite well belong to this category.

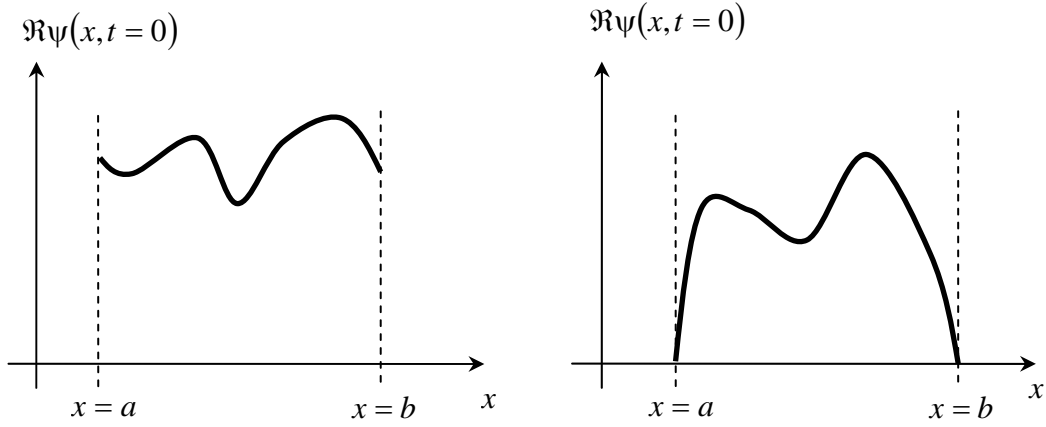

Figure 2: The real part of a bounded wavefunction at $t = 0$ with two singular points. On the left there is a discontinuity in the function and on the right the discontinuity is in its derivative.

In this case eq.2 should be replaced with

$$\varphi(x, t \geq 0) = \int_a^b \frac{dy}{2\sqrt{\pi i t}} \exp\left[i(x-y)^2/4t\right] \varphi(y, t=0) \tag{11}$$

and since $\int_a^b = \int_{-\infty}^b - \int_{-\infty}^a$ eq.6 should be written

$$\psi(x, t \geq 0) = \frac{1}{2} \exp\left(i \frac{(x-b)^2}{4t}\right) w\left(\frac{-i2t\partial/\partial y - (x-b)}{2(-it)^{1/2}}\right) \psi(y,0)\big|_{y=b} - \frac{1}{2} \exp\left(i \frac{(x-a)^2}{4t}\right) w\left(\frac{-i2t\partial/\partial y - (x-a)}{2(-it)^{1/2}}\right) \psi(y,0)\big|_{y=0} \tag{12}$$

and at short times

$$\psi(x, t \ll x^2) \sim \sqrt{\frac{it}{\pi}} \left\{ \frac{\exp\left(i(x-b)^2/4t\right)}{x-b} \psi(b,0) - \frac{\exp\left(i(x-a)^2/4t\right)}{x-a} \psi(a,0) \right\} \tag{13}$$

These two terms interfere, and the oscillating interference term appear in the probability density

$$|\psi(x,t)|^2 \sim \frac{t}{\pi} \left\{ \frac{|\psi(b,0)|^2}{(x-b)^2} + \frac{|\psi(a,0)|^2}{(x-a)^2} - 2\Re\left( \frac{\exp\left[i\left(x - \frac{a+b}{2}\right)\frac{a-b}{2t}\right]}{(x-b)(x-a)} \psi(b,0)\psi^*(a,0) \right) \right\} \tag{14}$$

Note that the propagating term propagates with respect to the center of the bounded region $(a+b)/2$ with velocity that depends on its width and decreases with time $v \sim (a-b)/2t$. The oscillations' wavelength is proportional to $t$ and inversely related to the initial width of the wavepacket. The third oscillating term is a manifestation of the interference between the two boundaries, and is totally absent in Moshinsky's ("single boundary") function.

When the initial wavefunction vanishes at the boundaries (right plot of fig.2) the propagation at short times goes like

$$\psi(x,t) \sim 4\frac{t^3}{\pi} \left\{ \frac{|\psi'(b,0)|^2}{(x-b)^4} + \frac{|\psi'(a,0)|^2}{(x-a)^4} - 2\Re\left( \frac{\exp\left[i\left(x - \frac{a+b}{2}\right)\frac{a-b}{2t}\right]}{(x-b)^2(x-a)^2} \psi'(b,0)\psi'^*(a,0) \right) \right\} \tag{15}$$

were the tags stand for spatial derivatives $\psi'(c,0) \equiv \partial\psi(y,0)/\partial y|_{y=c}$.

We would like to emphasize that these last equations suggest that at short times the wavefunction propagation depends only on the initial values of the wavefunction at the boundaries.

These results can easily be generalized to other cases with physical relevancy. For example, in the well known cases of the double slits, where the first slit boundaries are $x_1$ and $x_2$ and the second one's are $x_3$ and $x_4$ the wavefunction at short times can be written

$$\psi(x, t \ll x^2) \sim \sqrt{\frac{it}{\pi}} \sum_{j=1}^{4} (-1)^{j+1} \frac{\exp(i(x-x_j)^2/4t)}{x-x_j} \psi(x_j, 0). \tag{16}$$

Obviously, the interference terms include beats, which are reminiscent of the fact that frequency of the interference terms depends on the coordinate $x$.

As was said in the introduction, it is possible to question the validity of these conclusions in a physical, i.e., non singular, case. However, it can be easily shown (see sec. IV) that even when the singular discontinuity is replaced with a sharp but continuous transition, the above derivation is valid so long as the measurement time is not too short.

**III Numerical solutions.** To validate these results we solved the propagation integrals numerically. In fig.3 we present two arbitrary wavefunctions (one is a constant and the other is an oscillating function), which have the same values at the region's boundaries. Their initial probability density is plotted in the upper plot of fig.4. In the lower plot we present their values (multiplied by $x^2$ to eliminate the slow decay) after a certain time $\Delta t$ (note that physically the unit of time $\tau$ depends on the chosen unit of distance in the plot, say $\Delta x$, then, $\tau = 2m\Delta x^2/\hbar$).

Despite the differences between the two functions in the region's interior part, the propagation at short times, i.e, $t \ll x^2$, is almost identical (see the lower plot).

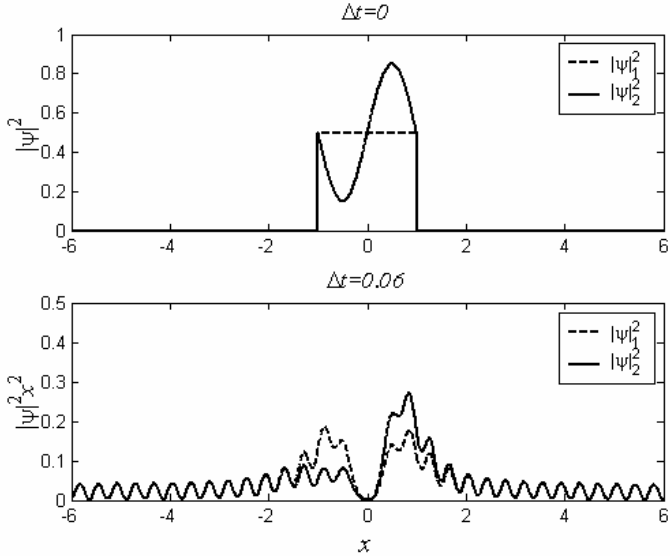

Figure 3: Short time propagation of two different bounded wavefunctions, which have initially the same values at the boundaries (upper plot). Despite the differences, the propagation is similar at short times, $t \ll x^2$ (lower plot).

In Fig. 4 we present a similar simulation but for a wave-packet that vanishes at the boundaries, i.e., the function is continuous there but its derivative is not. In that case we can see again that the short time propagation of two different functions depends only on its derivatives' value at the boundaries.

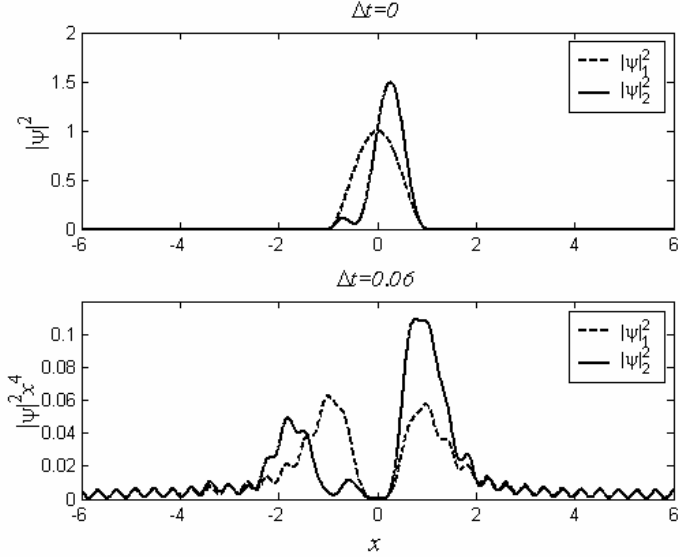

Figure 4: Short time propagation of two different bounded wavefunctions, which have initially the same derivative values at the boundaries (upper plot). Despite the differences, the propagation is similar at short times, $t \ll x^2$ (lower plot).

**IV Continuous boundaries.** In most physical problems sharp boundaries are continuous rather than singular, and in order to show that the universal behavior, which appears in eqs.9 and 10, is meaningful even in these continuous cases, we have to investigate the limit from the continuous case to the singular one. We choose two functions. One continuous

$$\psi_1(x, t=0) = [1 - \tanh(x/\xi)]/2 \tag{17A}$$

and the other is the singular step (Heavyside) function

$$\psi_2(x, t=0) = \theta(-x) \tag{17B}$$

where clearly, $\psi_2$ is the singular limit of $\psi_1$, i.e., $\psi_1(x) \xrightarrow{\xi \to 0} \psi_2(x)$.

At finite $t > 0$ (see, for example, ref.14)

$$\psi_1(x, t \geq 0) = \int_{-\infty}^{\infty} dk \left[ \frac{1}{2}\delta(k) + i\frac{\xi}{4}\operatorname{cosech}(\pi k \xi / 2) \right] \exp(ikx - ik^2 t) \tag{18A}$$

and

$$\psi_2(x, t \geq 0) = \int_{-\infty}^{\infty} dk \left[ \frac{1}{2}\delta(k) + \frac{i}{2\pi k} \right] \exp(ikx - ik^2 t). \tag{18B}$$

$\psi_2$ is, of course, a particular case of the Moshinsky function.

After expanding the cosech(x) function

$$\psi_1(x, t \geq 0) = \int_{-\infty}^{\infty} dk \left[ \frac{1}{2}\delta(k) + i\frac{1}{2\pi k} - i\frac{\pi}{48}\xi^2 k + O(\xi^4 k^3) \right] \exp(ikx - ik^2 t) \tag{19}$$

and the difference between the two functions in the limit $\xi \to 0$ is

$$\psi_1(x, t \geq 0) - \psi_2(x, t \geq 0) \sim -\frac{\pi}{48} \int_{-\infty}^{\infty} \left( ik\xi^2 + O(\xi^4 k^3) \right) dk \exp(ikx - ik^2 t) \tag{20}$$

One can see, that since the function cosech(x) is odd, each term is $\sim \xi^2 k^2$ smaller than its predecessor, and since each $k$ corresponds to an additional spatial derivative of $\exp(ix^2/4t)$, then it is clear that this series converges to a negligible value, only when $\xi x/t \ll 1$.

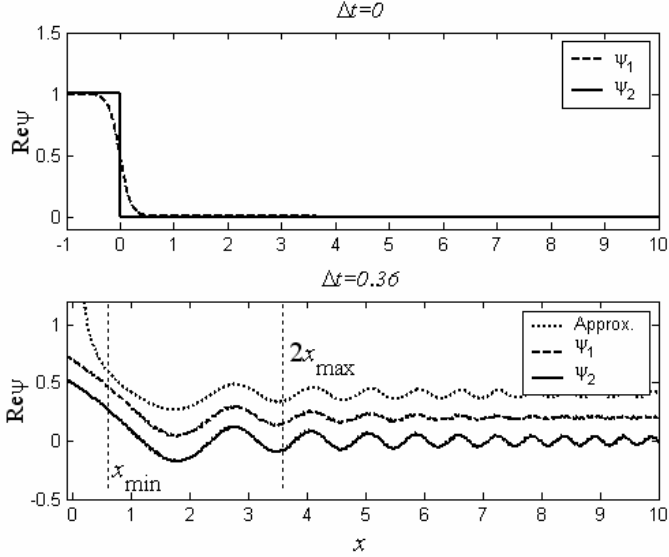

Figure 5: A comparison between two types of boundaries: the continuous $\psi_1(x,0) = [1 - \tanh(x/\xi)]/2$ ($\xi = 0.2$) and the singular step function $\psi_2(x,0) = \theta(-x)$. In the upper panel we show the initial condition. The lower panel presents the temporal propagation for the two functions and the approximation (eq.9). The three plots were artificially shifted by 0, 0.2 and 0.4 (for $\psi_2, \psi_1$ and the approx.) respectively to help the comparison. In this plot $x_{min} \equiv \sqrt{\Delta t}$ and $x_{max} \equiv \Delta t / \xi$.

In fig.5 we plot the temporal propagation of the two boundaries. As can be seen from this figure, in the regime $x \ll x_{min} \equiv \sqrt{t}$ the approximation (eq.9) is no longer valid. On the other hand, where $x \gg x_{max} \equiv t/\xi$ the oscillations of $\psi_1(x,t)$ decay relatively fast, and the function cannot be described well enough by the singular $\psi_2(x,t)$. Therefore, the approximation (eq.6) can depict the propagation of the continuous boundary $\psi_1(x,t)$ with a relatively high accuracy in the spatial regime $x_{min} \ll x \ll x_{max}$, which in principle can be arbitrarily large.

This section was added to show that the results of the previous sections are valid, at least partially, even in the continuous case; however, one can use the conclusions of the present section to validate experimentally how a certain wave-packet was initially localized. As a particular case, such an experiment may help in analyzing the effect of a quantum measurement on a wavepacket.

**V Experimental realization.** In experimental setups, which were designed to investigate atom-optics billiards [6, 7] with Rubidium atoms, it was relatively simple to construct an approximately $200\mu m$ width atom-trap, whose walls (boundaries) were about an order of magnitude smaller. These traps (billiards), which are practically made of light, can be turned off instantaneously. If we measure the Rb atoms density about ~1mm from the trap, we should expect to find the universality of eq.15 (since the initial wavefunction vanishes at the boundaries) within the time range of $20\sec \ll t \ll 2000\sec$. Since the effective potential inside the trap can be modified, one can validate this universality for different scenarios and initial conditions.

**VI Summary and conclusions.** In this paper we investigated the propagation of a wavepacket with sharp edges (boundaries). The paper includes three main results and conclusions:

A) The propagation of any localized (or singular) wavepacket can be fully described at any time by a formula that depends only on the wavefunction (and its derivatives) values at the boundaries.

B) At short times $t \ll 2mx^2/\hbar$ the wavefunction propagation depends only on the value of the wavefunction at the boundary, $\psi(x,t) \sim \sqrt{it/\pi} x^{-1} \exp(ix^2/4t)\psi(0,0)$ and if the wavefunction vanishes there, it depends only on its first derivative. Since these expressions have an oscillating phase that depends on the location $x$, an additional conclusion arises, that when there is more than one boundary an oscillating interference term (or even beats) appears at the probability density $|\psi|^2$.

C) The first two conclusions are valid even for continuous rather than singular functions, provided the spatial transitions at the boundaries are short enough. If the change scale at the boundary is $\xi$ then conclusions A and B are valid in the regime $2mx\xi/\hbar \ll t \ll 2mx^2/\hbar$. Thus, the region where the interference oscillations exist can be a measure to the initial wavefunction localization.